# Validation of GFS day-ahead solar irradiance forecasts in China

**Yue Zhang**[1,2], Yanbo Shen[3], Xiangao Xia[2,4], Guangyu Shi[1,2]

1. LASG, Institute of Atmospheric Physics, Chinese Academy of Sciences, Beijing, China
2. University of Chinese Academy of Sciences, Beijing, China
3. Public Meteorological Service Center, China Meteorological Administration, Beijing, China
4. LAGEO, Institute of Atmospheric Physics, Chinese Academy of Sciences, Beijing, China

**Abstract**

This paper provides a benchmark to evaluate operational day-ahead solar irradiance forecasts of Global Forecast System (GFS) for solar energy applications in China. First, GFS day-ahead solar irradiance forecasts are validated quantitatively with hourly observations at 17 first-class national solar monitoring stations and 1 Baseline Surface Radiation Network (BSRN) station all over China. Second, a hybrid forecast method based on Gradient Boosting (GB) and GFS product is proposed to improve forecasts accuracy. Both GFS forecasts and GB-based forecasts are compared with persistence forecasts. The results demonstrate persistence model is more accurate than GFS forecasts, and the hybrid method has the best performance. Besides, parameter optimization of direct-diffuse separation fails to reduce the errors of direct normal irradiance (DNI) forecasts.

**Nomenclature**

ACCESS
    Australian Community Climate and Earth-System Simulator

ANN
    Artificial Neural Network

ARM
    Atmospheric Radiation Measurement

BOM
    Bureau of Meteorology

BRL
    Boland–Ridley–Laurent

BSRN
    Baseline Surface Radiation Network

CCAM
    Conformal Cubic Atmospheric Model

CSP
    Concentrating Solar Power

CMA
    China Meteorological Administration

CPV
    Concentrator Photovoltaics

DHI
    Diffuse Horizontal Irradiance

DNI
    Direct Normal Irradiance

DSWRF
    Downward Short-Wave Radiation Flux

ECMWF
    European Centre for Medium-Range Weather Forecasts

GB

Gradient Boosting

GBRT

Gradient Boosted Regression Trees

GFS

Global Forecast System

GHI

Global Horizontal Irradiance

HARMONIE-AROME

HIRLAM-ALADIN Regional Meso-scale Operational NWP In Europe-Application of Research to Operations at Mesoscale

IEA SHC

International Energy Agency, Solar Heating and Cooling Programme

IFS

Integrated Forecasting System

JMA

Japan Meteorological Agency

JMA MSM

Japan Meteorological Agency, Meso-Scale Model

LW

Longwave

McICA

Monte-Carlo Independent Column Approximation

MESoR

Management and exploitation of solar resource knowledge

ML

Machine Learning

MOS

Model Output Statistics

MSM

Meteorological Mesoscale Model

NAM

    North American Mesoscale Model

NCEP

    National Centers for Environmental Prediction

NDFD

    National Digital Forecast Database

NOMADS

    NOAA Operational Model Archive and Distribution System

NREL

    National Renewable Energy Laboratory

NWP

    Numerical Weather Prediction

PRES

    Surface Pressure

PV

    Photovoltaics

PWAT

    Precipitable Water

QC

    Quality Control

RH

    Relative Humidity at 2m

rMAE

    relative Mean Absolute Error

rMBE

    relative mean bias error

RMSE

    root mean squared error

rRMSE

    relative root mean squared error

RRTMG

    Rapid Radiative Transfer Model for GCMs

RTM

    Radiative Transfer Model

SERI

    Solar Energy Research Institute

SUNSD

    Sunshine Duration

SURFRAD

    Surface Radiation Budget Network

SW

    Shortwave

TCDC

    Total Cloud Cover

TMP

    Surface Temperature

UTC

    Coordinated Universal Time

WRF

    Weather Research and Forecasting Model

XGBoost

    Extreme Gradient Boosting

# 1. Objective & Background

Solar energy is a clean and renewable energy source. As a mitigation solution to climate change, air pollution, and fossil fuel shortage, solar power has experienced strong annual growth worldwide especially in China over the last decade. Given the intermittence and volatility of surface solar irradiance, power generation of a solar power plant, i.e., Photovoltaics (PV), Concentrator Photovoltaics (CPV) and Concentrating Solar Power (CSP), heavily depends on meteorological conditions, therefore complicating the schedule and operation of Transmission System Operator (TSO). To reduce the impact of large-scale solar integration on electricity grid, day-ahead power generation of a solar power plant should be forecasted and reported to TSO in advance. Admittedly accurate surface solar irradiance forecast is the prerequisite to provide reliable solar power forecasting.

There are many approaches to forecast surface solar irradiance in different time scales (Pelland et al., 2013). Numerical Weather Prediction (NWP) is a powerful computer tool which uses a set of governing equations to describe the flow of fluids like the atmosphere or the ocean, and forecasts the future state of the atmosphere or the ocean by ingesting current meteorological observations through a procedure known as data assimilation. For day-ahead time horizon, NWP is the most common method to predict surface solar irradiance. Because weather forecast is complex and computationally extensive, NWP models always run on supercomputers and are operated by national-level weather centers, e.g., European Centre for Medium-Range Weather Forecasts (ECMWF), National Centers for Environmental Prediction (NCEP), The Bureau of Meteorology (BOM), Japan Meteorological Agency (JMA), and China Meteorological Administration (CMA), etc. Operational NWP model products, which cover global or regional ranges, can directly be used in relevant application such as solar and wind energy, and also can serve as the initial or boundary conditions of a customized mesoscale model like Weather Research and Forecasting Model (WRF), which has worldwide users in energy meteorology community.

Since the performance of an operational NWP model mainly relies on climate and terrain characteristics of interest areas, evaluating the accuracy of surface solar irradiance forecast by an operational NWP model in specific region is essential to all stakeholders of solar power forecasting, involving solar power plants, TSOs, financial investors, weather bureaus, and third-party forecast service providers. Motivated by this demand, the performance of operational NWP models had been extensively validated with ground-based measurements (Gala et al., 2016; Gregory and Rikus, 2016; Huang et al., 2018; Huang and Thatcher, 2017; Joshi et al., 2019; Landelius et al., 2018; Lima et al., 2016; Lopes et al., 2018; Lorenz et al., 2009; Lorenz et al., 2016; Mathiesen and Kleissl, 2011; Ohtake et al., 2013; Ohtake et al., 2019; Perez et al., 2010; Perez et al., 2013; Remund et al., 2008; Troccoli and Morcrette, 2014; Verbois et al., 2018).

Within the framework of IEA SHC (International Energy Agency, Solar Heating and Cooling Programme) Task 36, Remund et al. (2008) evaluated hourly global horizontal irradiance (GHI) forecasts of ECMWF, National Digital Forecast Database (NDFD) and WRF against Surface Radiation Budget Network (SURFRAD) ground observation in the US, and found that ECMWF has the best performance. As a complement, Lorenz et al. (2009) validated 7 models from IEA SHC Task 36 participants with hourly ground-based irradiance measurements at 24 stations in Germany, Switzerland, Austria, and Spain. Both benchmarking studies presented a strong dependency of forecast uncertainties on local climatic conditions. IEA SHC Task 46, which continued the work of IEA SHC Task 36 on model inter-comparison, compared hourly GHI forecasts based on 5 NWP models and 2 Model Output Statistics (MOS) methods against ground truth up to 61 sites in Denmark, Germany and Switzerland, and found that spatial and temporal averaging to a certain extent can improve forecast accuracy (Lorenz et al., 2016). Mathiesen and Kleissl (2011) evaluated intra-day GHI forecasts of ECMFW, Global Forecast System (GFS) and North American Mesoscale Model (NAM) over the US with SURFRAD hourly observations, and analyzed the impact of MOS on predictions. It demonstrated that for direct model output, ECMWF outperforms other models and GFS performs best after MOS correction. For day-ahead and multi-day GHI forecasts over the US, Perez et al. (2011) evaluated the performance of 7 global and mesoscale models with SURFRAD dataset, which presented that global models' prediction ability is better than mesoscale ones, and simple averaging of multi-model outperforms individual model. On the basis of Perez et al. (2011), Perez et al. (2013) extended the spatial range for validation of NWP solar irradiance forecasts to Canada and Europe including Germany, Switzerland, Austria and Spain, in which 10 global, multiscale and mesoscale models were evaluated against ground data at 34 sites, and the results were in line with previous work over the US. Ohtake et al. (2013) investigated the characteristics of day-ahead GHI forecasts by JMA

Meteorological Mesoscale Model (MSM) model with ground measurement at 5 sites over the Kanto region, and proposed that refining the representation of specific clouds in the model would improve forecast accuracy. Huang and Thatcher (2017) compared day-ahead GHI forecasts by GFS and Conformal Cubic Atmospheric Model (CCAM) against 9 sites operated by BOM over Australia, and presented spatial averaging can improve forecast accuracy and its optimal scale is determined by on-site climatic conditions. Verbois et al. (2018) conducted a comparison concerning day-ahead GHI forecasts between WRF and GFS model against ground truth at 25 meteorological stations over Singapore, in which GFS outperforms WRF with distinct configurations during some of tested months, and proposed that for specific applications implementing GFS mean forecasts are more reasonable than running WRF at high resolution which is computationally expensive.

NWP models calculate the shortwave (SW) radiation with radiative transfer model (RTM), which is tailored and outputs global radiation merely. To retrieve direct and diffuse irradiance forecasts, direct-diffuse separation was used in early studies. Later a few NWP models upgraded the radiative transfer scheme in order to output direct and diffuse components. For example, ECMWF upgraded its IFS model and added direct horizontal radiation component since Cy37r2 to deliver better forecast product to CSP industry. WRF also released an augmented version designed for solar application WRF-Solar (Jimenez and Hacker, 2016). These configurations make benchmark of direct and diffuse solar irradiance forecasts be possible.

Breitkreuz et al. (2009) evaluated the performance of short-term direct normal irradiance (DNI) forecasts by ECMWF against ground-based measurements at 121 sites over Europe and the Mediterranean area. It was found that under clear-sky conditions aerosols dominate the accuracy of DNI forecasts, and NWP model struggles to predict DNI steadily for cloudy sky. Since ECMWF irradiance predictions have no direct normal component, a direct-diffuse separation model is used to retrieve DNI forecasts (Skartveit et al., 1998). Troccoli and Morcrette (2014) evaluated the forecast performance of global and direct horizontal irradiance by two versions of ECMWF model against observations at 4 BOM sites distributed over Australia, and the results showed that forecast errors strongly depend on sky conditions and on-site climates. Following the work by Troccoli and Morcrette (2014), Gregory and Rikus (2016) validated GHI, DNI and diffuse horizontal irradiance (DHI) forecasts by BOM's new generation operational model

Australian Community Climate and Earth-System Simulator (ACCESS) with data at 8 stations over Australia. It presented that large errors for DNI and DHI forecasts result from insufficient scattering of direct irradiance through clouds in radiative transfer scheme. Schroedter-Homscheidt et al. (2017) found that introducing direct irradiance in ECMWF forecast variables inventory does not result in an overall performance improvement of DNI predictions in comparison with conventional direct-diffuse separation. Landelius et al. (2018) compared GHI and DNI forecasts by deterministic and ensemble ECMWF with HARMONIE-AROME ensemble predictions using 1 site in Sweden, and the results presented that control member of HARMONIE-AROME ensemble outperforms ECMWF deterministic forecast. Lopes et al. (2018) evaluated 24 h GHI and DNI forecasts by ECMWF over Portugal against ground measurements at 4 stations and found that ECMWF predicts GHI well while DNI is more difficult to predict due to aerosol and cloud representation in the model. Joshi et al. (2019) validated GHI, direct horizontal irradiance and DHI forecasts by two versions of BOM ACCESS models with high and coarse resolution against ground data at 11 stations across Australia and found that forecasts of direct and diffuse components are less accurate than GHI.

Many aforementioned cases demonstrated that operational global models have superior performance than mesoscale models. A possible explanation is that higher resolution of a mesoscale model results in spatial and temporal offsets of the cloud field which further decrease the accuracy of surface solar irradiance forecasts (Zack, 2012).

Along with the progress of artificial intelligence, many hybrid methods for solar irradiance forecast which combine NWP product and machine learning (ML) algorithms were proposed (Cornaro et al., 2015; Gala et al., 2016; Lauret et al., 2014; Marquez and Coimbra, 2011; Pereira et al., 2019). Marquez and Coimbra (2011) proposed a hybrid GHI and DNI forecasts method in which NDFD meteorological variables and solar geometric parameters are considered to be the inputs of Artificial Neural Network (ANN) algorithm, and validation results presented the cosine of solar zenith angle, normalized hour angle, cloud cover, precipitation probability, maximum and minimum temperatures are the most important features for ANN-based predictions. Lauret et al. (2014) presented an ANN-based post-processing model for day-ahead WRF GHI forecasts in which the inputs include clearness index and solar zenith angle. Validation results at Reunion Island demonstrated that this correction model significantly reduces the bias of one-point forecasts and spatial averaged forecasts. Cornaro et al. (2015) proposed a hybrid model combining day-ahead ECMWF GHI forecasts and ANN algorithm in which master optimization process is used to search the optimal neuron number as well as ANN ensemble. Benchmark accuracy at Rome showed that this model improves about 30% root mean squared error (RMSE) in comparison with persistence model. Gala et al. (2016) used three ML algorithms to develop hybrid forecasting models which ingest ECMWF radiation and cloud cover variables as inputs, and evaluated the performance of hybrid methods against the observations at 7 sites over Spain. It presented that forecasts based on support vector regression outperform other methods. Pereira et al. (2019) presented an ANN-based correction model for ECMWF GHI forecasts in which the inputs of the algorithm consist of ECMWF variables and predicted GHI with a clear-sky model. Validation results against 4 sites over Portugal showed that this hybrid method improves the GHI forecasts successfully.

Literature review presents that operational NWP models for solar irradiance forecasts are seldom evaluated with radiometric observations over China, which has the largest solar power capacity in the world. To the best of the authors' knowledge, GFS is the most widely used operational NWP product by energy meteorology community in China because of its accessibility. Hence, this study focuses on benchmark of day-ahead GFS irradiance forecasts over China. GFS day-ahead GHI, DNI and DHI forecasts are validated against a nationwide network of 17 radiometric stations and Baseline Surface Radiation Network (BSRN) Xianghe station. BRL, a direct-diffuse separation model, is also evaluated with same ground truth before it is used to derive DNI and DHI forecasts. Besides, a hybrid method for GHI, DNI and DHI forecasts, based on GFS product and machine learning algorithm Gradient Boosting (GB), is developed and validated with observations at BSRN Xianghe station. Later, GFS and hybrid method are compared with a persistence model. The rest of this article is organized as follows. Section 2 presents GFS forecast product and ground-based measurements used for validation. Section 3 describes quality control procedure for radiometric data, direct-diffuse separation method, GB algorithm, persistence model and error metrics. Section 4 discusses the validation results. Finally, section 5 summarizes the conclusions.

**2. Numerical weather prediction and ground observation**

2.1 Global Forecast System (GFS)

GFS is a global weather forecast model maintained and operated by NCEP. GFS radiation scheme solves the SW and longwave (LW) radiative fluxes based on both modified and optimized versions of the Rapid Radiative Transfer Model for GCMs (RRTMG) (Clough et al., 2005; Iacono et al., 2000; Iacono et al., 2008; Mlawer et al., 1997). SW radiation scheme consists of 14 broad spectral bands, in which absorption effects from ozone, water vapor, carbon dioxide, methane, nitrous oxide, and oxygen are considered. Cloud condensate path, effective radius for water and ice, water clouds, and ice clouds are also solved to represent cloud-radiation interactions (Fu, 1996; Hu and Stamnes, 1993). Subgrid variability in multi-level clouds is represented statistically using Monte-Carlo Independent Column Approximation (McICA), and the vertical relationship is calculated with maximum-random cloud overlap method. SW scheme calculates aerosol-radiation interactions among various aerosol components in each spectral band with a 5-degree climatological aerosol in troposphere, and a volcanic aerosol parameterization in stratosphere separately. SW albedo is parameterized with the types of surface vegetation (Hou et al., 2002), and the dependency of snow-free land surface albedo on solar zenith angle is modified (Yang et al., 2008). More information on GFS radiation scheme can be found in (https://vlab.ncep.noaa.gov/web/gfs/documentation).

GFS forecast, covers dozens of atmospheric variables in a grid with a base horizontal resolution of 25 kilometers (T1534), is published at 00, 06, 12 and 18 UTC (Coordinated Universal Time) with 3 h timestep up to 240 h ahead. Post-processed grid data with distinct resolutions are available publicly to online users. To fit the application scenario of domestic electricity power industry in China, GFS 004 with spatial resolution of 50 kilometers issued at 18 UTC cycle is used in this retrospective forecast experiment. GFS archive datasets were retrieved from NOAA Operational Model Archive and Distribution System (NOMADS) website (http://nomads.ncep.noaa.gov/) and pre-processed with rNOMADS, an open source R package (Bowman and Lees, 2015). Required GFS variables in this study are downward short-wave radiation flux (DSWRF), surface pressure (PRES), surface temperature (TMP), relative humidity at 2m (RH), sunshine duration (SUNSD), precipitable water (PWAT), and total cloud cover (TCDC) (Table 1). DSWRF spans from Jan 1st, 2014 to Aug 31st, 2016, and the others cover Jan 1st, 2014 to Dec 31st, 2015. Considering distinct temporal resolution, above variables except

DSWRF are linearly interpolated to 1 h with pre-processing. Given the relationship between sun position and time, DSWRF is interpolated to 1 h with clearness index.

Table 1. Overview of required GFS variables

| No. | Abbrevation | Level | Parameter | Forecast valid | Unit |
|---|---|---|---|---|---|
| 1 | DSWRF | surface | downward short-wave radiation flux | 0-3 hour ave<br>0-6 hour ave | W/m$^2$ |
| 2 | PRES | surface | pressure | 6 hour fcst | Pa |
| 3 | TMP | surface | temperature | 6 hour fcst | K |
| 4 | RH | 2m above ground | relative humidity | 6 hour fcst | % |
| 5 | SUNSD | surface | sunshine duration | 6 hour fcst | s |
| 6 | PWAT | entire atmosphere | precipitable water | 6 hour fcst | kg/m$^2$ |
| 7 | TCDC | entire atmosphere | total cloud cover | 0-6 hour ave | % |

2.2 Solar irradiance measurements

China has the largest installed solar power capacity in the world. Weather stations for radiometric and meteorological measurements had been widespread deployed with the construction of solar power plants. However, the measurements are not always available unfortunately, and in the cases when they are, due to instrument malfunction and especially poor maintenance the data suffers from missing values and quality issue commonly. Therefore, data from such radiometric stations are inadequate for model validation.

2.2.1 CMA First-class stations

China Meteorological Administration (CMA) operates an extensive national solar radiation monitoring network, comprising 98 radiometric stations. The stations are categorized into three classes, as first-class, second-class and third-class stations based on instrument type. 17 first-class stations out of 98 stations measure GHI, DNI and DHI independently with thermopile radiometers, thus offering the possibility to validate the forecasts of three components by NWP models. Radiometers in CMA network had been updated several times since 1960s. After 1990, DFY-4 pyranometer was used to gauge GHI and DHI while DNI was measured with DFY-3 pyrheliometer. Since 2003 CMA has measured GHI and DHI with TBQ-2-B pyranometer, and DNI has been collected with TBS-2-B pyrheliometer. Detailed information about instruments at first-class stations are presented in Table 2. In this work, hourly solar irradiance datasets of 17 first-class sites between Jan 1st, 2015 and Dec 31st, 2016 were retrieved from National Meteorological Information Center. All the measurements are subject to quality control described in section 3.1 before the use (Hoyer-Klick et al., 2008).

Table 2. Specifications of solar monitoring instruments at CMA first-class stations

|  | GHI | DNI | DHI |
|---|---|---|---|
| Model | TBQ-2-B | TBS-2-B | TBQ-2-B |
| Manufacturer | Huatron-HSC | Huatron-HSC | Huatron-HSC |
| Classification | First class | First class | First class |
| Spectral range | 300-3000 nm | 300-3000 nm | 300-3000 nm |
| Field of view | 180° | 4° | 180° |
| Sensitivity | 7-14 μV/W/m$^2$ | 7-14 μV/W/m$^2$ | 7-14 μV/W/m$^2$ |
| Response time (99%) | ≤ 35 s | ≤ 25 s | ≤ 35 s |

2.2.2 BSRN Xianghe station

Xianghe station (Latitude: 39.754°N, Longitude: 116.962°E, Elevation: 36 m), operated by Institute of Atmospheric Physics at Chinese Academy of Sciences, is the only BSRN station in China mainland. Surface solar irradiance had been measured continuously since September 2004 (with an 8-month interruption from July 1st, 2012 to February 28th, 2013). GHI was measured by Kipp & Zonen CM21/CM11 secondary standard pyranometer when DNI and DHI were gauged respectively with Eppley Normal Incidence pyrheliometer and Black & White pyranometer, which were mounted on an EKO STR-22 sun tracker (Xia et al., 2007). Specifications of instruments are listed in Table 3. Solar irradiance was sampled every second and data logger only stored 1-min average. In this study, 1-min averages between Oct 2nd, 2004 and Nov 1st, 2015 are aggregated to hourly averages, after removing the samples which fail to pass quality control (Hoyer-Klick et al., 2008).

Among 18 radiometric stations, Xianghe is the only research-class site and has much more historical observations. Hence, Xianghe data is also used to study parameter optimization of direct-diffuse separation model and develop hybrid forecasting model. A description of all 18 stations is shown in Table 4, and their geographical location is depicted in Fig. 1.

Table 3. Specifications of solar monitoring instruments at Xianghe

|  | GHI | DNI | DHI |
|---|---|---|---|
| Model | CM11 / CM21 | NIP | Black & White |
| Manufacturer | Kipp & Zonen | Eppley Lab | Eppley Lab |
| Classification | Secondary standard | Secondary standard | First class |
| Spectral range (50%) | 305-2800 nm | 250-3000 nm | 295-2800 nm |
| Field of view | 180° | 5° | 180° |
| Sensitivity | 4-6 μV/W/m$^2$ / 7-17 μV/W/m$^2$ | 8 μV/W/m$^2$ | 8 μV/W/m$^2$ |
| Response time (95%) | 12 s / 5 s | 5 s | 30 s |
| Uncertainty (hourly) | 3% / 2% | 1% | 3-5% |

Table 4. Details on 18 ground stations, including site name, id, latitude and longitude in degrees and elevation in meters above mean sea level.

| Station | ID | Lat (degree) | Long (degree) | Elevation (m) |
|---|---|---|---|---|
| Mohe | 50136 | 52.97 | 122.52 | 438.5 |
| Harbin | 50953 | 45.93 | 126.57 | 118.3 |
| Urumqi | 51463 | 43.78 | 87.65 | 935 |
| Kashi | 51709 | 39.48 | 75.75 | 1385.6 |
| Ejinaqi | 52267 | 41.95 | 101.07 | 940.5 |
| Geermu | 52818 | 36.42 | 94.92 | 2807.6 |
| Yuzhong | 52983 | 35.87 | 104.15 | 1874.4 |
| Shenyang | 54342 | 41.73 | 123.52 | 49 |
| Beijing | 54511 | 39.80 | 116.47 | 31.3 |
| Lhasa | 55591 | 29.67 | 91.13 | 3648.9 |
| Chengdu | 56187 | 30.75 | 103.87 | 547.7 |
| Kunming | 56778 | 25.00 | 102.65 | 1888.1 |
| Zhengzhou | 57083 | 34.72 | 113.65 | 110.4 |
| Wuhan | 57494 | 30.60 | 114.05 | 23.6 |
| Baoshan | 58362 | 31.40 | 121.45 | 5.5 |
| Guangzhou | 59287 | 23.22 | 113.48 | 70.7 |
| Sanya | 59948 | 18.22 | 109.58 | 419.4 |
| Xianghe | XH | 39.75 | 116.96 | 36 |

## 3. Method

3.1 Quality control of ground-based measurements

Many quality control (QC) methods are available to flag the erroneous samples of surface solar irradiance, e.g., NREL SERI (National Renewable Energy Laboratory, Solar Energy Research Institute) QC procedure (Maxwell et al., 1993), BSRN Global Network recommended QC tests (Long and Dutton, 2002), ARM (Atmospheric Radiation Measurement) QCRad methodology (Long and Shi, 2008), and MESoR (Management and exploitation of solar resource knowledge) QC procedure (Hoyer-Klick et al., 2008) etc. For present study, data samples fail to follow below conditions are flagged and removed (Eq. (1-7)).

$$-4 < GHI < 1.5 \times I_0 \times sin^{1.2}(\alpha) + 100 \tag{1}$$

$$-4 < DNI < I_0 \tag{2}$$

$$-4 < DHI < 0.95 \times I_0 \times sin^{1.2}(\alpha) + 50 \tag{3}$$

$DNI \times sin(\alpha) + DHI > 50$ W/m², $\alpha > 15^o$

$$1 - 8\% \leq \frac{GHI}{DNI \times sin(\alpha) + DHI} \leq 1 + 8\% \tag{4}$$

$DNI \times sin(\alpha) + DHI > 50$ W/m², $-3^o < \alpha < 15^o$

$$1 - 15\% \leq \frac{GHI}{DNI \times sin(\alpha) + DHI} \leq 1 + 15\% \tag{5}$$

$GHI > 50$ W/m², $\alpha > 15^o$

$$\frac{DHI}{GHI} < 1.05 \tag{6}$$

$GHI > 50$ W/m², $-3^o < \alpha < 15^o$

$$\frac{DHI}{GHI} < 1.10 \tag{7}$$

Additionally, data samples with solar elevation angle less than 5° are removed. Sun position parameters and extraterrestrial irradiance, which are frequently being referred to in this work, are calculated with the method described in (Goswami et al., 2000).

3.2 Direct-diffuse separation

Direct-diffuse separation aims to estimate direct and diffuse components when only global irradiance is available. As no direct and diffuse irradiance are included in parameter inventory of GFS forecast product, day-ahead DNI and DHI forecasts have to be derived with DSWRF. According to literature review, more than 100 direct-diffuse separation models were published since 1960s (Gueymard and Ruiz-Arias, 2016). Except global irradiance, many models also require other variables which are not always measured alongside solar irradiance components as inputs, and in those cases the applicability of a model is ultimately limited. Ridley et al., (2010) proposed an easy-to-use logistic model Boland–Ridley–Laurent (BRL) in which all predictors can be calculated through sun position parameters, extraterrestrial irradiance and observed GHI, and indicated that BRL performs better than other classic models in both hemispheres. As an empirical model, BRL's coefficients reflect on-site climatic pattern. Adjusting the coefficients of BRL to ground data at targeted area can improve decomposition accuracy. Lemos et al., (2017) adjusted BRL to local climate in Brazil by tuning model coefficients with ground data, and demonstrated that on hourly basis optimized BRL performs overall better than the original model proposed by (Ridley et al., 2010). Because of its usability, good performance and tunability, BRL is chosen for the present study (Eq. (8-9)).

$$d = \frac{1}{1+e^{A_1+A_2 \cdot k_t + A_3 \cdot AST + A_4 \cdot \alpha + A_5 \cdot K_t + A_6 \cdot \psi}} \tag{8}$$

$$DNI = \frac{GHI - GHI \cdot d}{\sin \alpha} \tag{9}$$

where $d$ is diffuse fraction, $k_t$ is hourly clearness index, $AST$ is apparent solar time, $\alpha$ is solar elevation angle, $K_t$ is daily clearness index, $\psi$ is persistence defined in (Ridley et al., 2010), and $A_1$, $A_2$, $A_3$, $A_4$, $A_5$, $A_6$ are empirical coefficients (see Table 5). After obtaining diffuse fraction of GHI with Eq. (8), DNI and DHI can be estimated simply based on trigonometric rule (Eq. (9)).

Table 5. Empirical coefficients of BRL. BRL's coefficients are estimated with ground-based measurements at 7 sites worldwide (Ridley et al., 2010). Coefficients of tuned BRL in this study are calculated with observation at Xianghe between Jan 1st, 2005 and Dec 31st, 2006.

| Model | Empirical coefficients | | | | | |
|---|---|---|---|---|---|---|
| | A1 | A2 | A3 | A4 | A5 | A6 |
| BRL | -5.38 | 6.63 | 0.006 | -0.007 | 1.75 | 1.31 |
| Tuned BRL | -6.36 | 8.04 | -0.109 | 0.003 | 3.65 | 0.2 |

Diffuse fraction would increase under clear-sky conditions when GHI exceeds its theoretical clear sky value, and data spread plot shows an upward climb at higher $k_t$ which is termed as cloud enhancement event (Engerer, 2015). BRL's sigmoid curve cannot capture cloud enhancement events occurred in the analysis of instantaneous irradiance data, e.g., 1-min data (Starke et al., 2018). However, cloud enhancement events are rare at hourly time scale, and transient effects of irradiance are negligible. Therefore, choosing BRL is reasonable in present study.

3.3 Gradient boosting (GB)

Gradient boosting (GB) is an ensemble method which converts weak models to a strong model using the gradients of loss function (Friedman, 2001). With decision trees as weak model, strong model is built by adding new weak models sequentially in which each of new decision trees is trained on the residuals from the preceding ensembles. Despite inaccurate weak models, the performance of the ensembles can be boosted significantly. Besides, a benefit of solving regression with GB is that the relative influence of each input on the ensemble can be measured, called feature importance.

GB has been widely used by the winners of Kaggle competitions (www.kaggle.com) and data scientists in industry. Although GB is less dominant in comparison with ANN and other ML algorithms for applications in solar resource community, a few cases can be found (Aler et al., 2017; Gagne II et al., 2017; Gala et al., 2016; Pedro et al., 2018; Persson et al., 2017). Gala et al. (2016) employed GB to predict global radiation using ECMWF meteorological variables as inputs. Aler et al. (2017) used Extreme Gradient Boosting (XGBoost) to partition direct and diffuse components with global solar irradiance through indirectly combining the predictions of up to 140 direct-diffuse separation models, or directly considering the variables of the existing models as inputs. Gagne II et al. (2017) built a hybrid GHI forecasts model with Gradient Boosted Regression Trees (GBRT) which takes GFS meteorological variables as inputs. Persson et al. (2017) presented a hybrid PV power forecasting approach which takes historical power data and meteorological variables by JMA MSM as inputs based on GBRT. Pedro et al. (2018) proposed an XGBoost-based method to forecast intra-hour GHI and DNI based on measured solar irradiance and sky images.

In this study, an optimized distributed GB python library, XGBoost, is implemented (Chen and Guestrin, 2016).

3.4 Persistence forecast

To benchmark the forecasts by both GFS and XGBoost-based approach, persistence forecast is used as a baseline model which assumes sky conditions do not change from previous day to current day, and only the sun's movement is considered. In Eq. (10), $GHI_{fcst}$ is hourly GHI forecasts, $K_{tp}$ and $H_{0c}$ are respectively daily average clearness index in previous day and hourly extraterrestrial horizontal irradiance in current day. Considering DNI and DHI forecasts can hardly be calculated with $K_{tp}$, it assumes that the performance of corresponding persistence models is comparable to GHI model at identical site.

$$GHI_{fcst} = K_{tp} \cdot H_{0c} \tag{10}$$

3.5 Statistical indictors

To quantitatively evaluate overall performance of models and algorithm in this paper, three conventional statistical indicators are used, and all expressed in percent of observed averages of corresponding irradiance components, i.e., relative mean absolute error (rMAE), relative root mean squared error (rRMSE) and relative mean bias error (rMBE) (Eq. (11-13)).

$$rMAE = \frac{1}{n}\frac{1}{y}\sum_{i=1}^{n}|y_i' - y_i| \tag{11}$$

$$rRMSE = \frac{1}{y}\sqrt{\frac{1}{n}\sum_{i=1}^{n}(y_i' - y_i)^2} \tag{12}$$

$$rMBE = \frac{1}{n}\frac{1}{y}\sum_{i=1}^{n}(y_i' - y_i) \tag{13}$$

where $y_i'$ and $y_i$ are the estimated and measured irradiance, $y$ is mean value of measured irradiance, $n$ is the number of samples. For rMAE, rRMSE and absolute value of rMBE, the lower the values of error metrics are, the better the performance is.

**4. Results**

4.1 Validation of BRL

As mentioned above, GFS model has no variables concerning DNI and DHI forecasts. GFS day-ahead DNI and DHI forecasts are calculated with DSWRF variable and direct-diffuse separation model BRL. Therefore, it is necessary to independently quantify the performance of BRL model prior to validation of DNI and DHI forecasts. In Section 4.1.1, original BRL is applied to individual stations. Then a graphical comparison for model fit overlaid against observed data is illustrated associated with error statistics, which permits inspecting the goodness of fit for BRL. In Section 4.1.2, to evaluate the influence of parameter optimization on DNI and DHI estimation, BRL is tuned with observation at Xianghe, and the results generated by original BRL and tuned BRL are intercompared.

4.1.1 BRL performance

Data quality at 18 radiometric stations varies from one to another, and data availability is diverse (Table 6). Fig. 2 shows qualitatively the relationships between hourly clearness index $k_t$ and diffuse fraction $d$ for ground-based measurements at 18 stations. The logistic curve captures the spread and variation of the data at most locations except Urumqi, Kashi, Ejinaqi, Shenyang and Chengdu in which $k_t$-$d$ charts do not present good fit over the observation that may result from data availability. Besides, the fitted curves at Urumqi, Ejinaqi, Shenyang and Chengdu are located above the middle of data spread, which indicate overestimated diffuse fraction. The errors are depicted quantitatively in Table 7. For DNI estimation, maximal errors appear at Kashi in which rMAE, rRMSE and rMBE are 52.38%, 87.35% and 41.91%. For DHI estimation, error maximum occurs at Ejinaqi in which rMAE, rRMSE and rMBE are 88.48%, 105.50% and 82.34%. Like Ejinaqi, BRL overestimates DHI at Urumqi, Shenyang and Chengdu in which rMBE are 10.09%, 18.29% and 13.08% respectively.

Table 6. Data availability after quality control at 18 ground stations. Solar irradiance measurements at 17 CMA first-class stations span from Jan 1$^{st}$, 2015 to Dec 31$^{st}$, 2016. Observed solar irradiance at Xianghe covers Oct 2$^{nd}$, 2004 to Nov 1$^{st}$, 2015 (with an 8-month interruption from July 1$^{st}$, 2012 to February 28$^{th}$, 2013).

| Station | ID | Lat (degree) | Long (degree) | Data samples |
|---|---|---|---|---|
| Mohe | 50136 | 52.97 | 122.52 | 2114 |
| Harbin | 50953 | 45.93 | 126.57 | 3129 |
| Urumqi | 51463 | 43.78 | 87.65 | 553 |
| Kashi | 51709 | 39.48 | 75.75 | 1968 |
| Ejinaqi | 52267 | 41.95 | 101.07 | 1340 |
| Geermu | 52818 | 36.42 | 94.92 | 3009 |
| Yuzhong | 52983 | 35.87 | 104.15 | 4445 |
| Shenyang | 54342 | 41.73 | 123.52 | 1363 |
| Beijing | 54511 | 39.8 | 116.47 | 5591 |
| Lhasa | 55591 | 29.67 | 91.13 | 2473 |
| Chengdu | 56187 | 30.75 | 103.87 | 1028 |
| Kunming | 56778 | 25 | 102.65 | 2461 |
| Zhengzhou | 57083 | 34.72 | 113.65 | 6729 |
| Wuhan | 57494 | 30.6 | 114.05 | 5689 |
| Baoshan | 58362 | 31.4 | 121.45 | 5806 |
| Guangzhou | 59287 | 23.22 | 113.48 | 6682 |
| Sanya | 59948 | 18.22 | 109.58 | 5514 |
| Xianghe | XH | 39.75 | 116.96 | 28471 |

Table 7. Error statistics of DNI and DHI modeling with BRL at 18 ground stations (unit: %). Solar irradiance measurements at 17 CMA first-class stations span from Jan 1st, 2015 to Dec 31st, 2016. Observed solar irradiance at Xianghe covers Jan 1st, 2007 to Dec 31st, 2011.

|  | DNI | | | DHI | | |
|---|---|---|---|---|---|---|
| Station | rMAE | rRMSE | rMBE | rMAE | rRMSE | rMBE |
| Mohe | 44.68 | 79.40 | 21.07 | 34.15 | 52.96 | -8.99 |
| Harbin | 34.32 | 55.97 | 7.00 | 24.25 | 36.32 | 0.19 |
| Urumqi | 43.84 | 56.89 | 2.76 | 80.16 | 95.46 | 10.09 |
| Kashi | 52.38 | 87.35 | 41.91 | 28.40 | 46.75 | -21.37 |
| Ejinaqi | 31.63 | 38.49 | -27.04 | 88.48 | 105.50 | 82.34 |
| Geermu | 31.62 | 49.64 | 18.32 | 34.16 | 49.57 | -19.40 |
| Yuzhong | 27.91 | 42.42 | 7.83 | 22.69 | 33.89 | -6.24 |
| Shenyang | 33.72 | 48.07 | -15.44 | 41.96 | 56.44 | 18.29 |
| Beijing | 25.99 | 35.17 | 7.34 | 22.63 | 30.92 | -4.96 |
| Lhasa | 40.30 | 61.67 | 28.94 | 43.90 | 63.84 | -29.96 |
| Chengdu | 38.66 | 80.14 | -12.44 | 16.60 | 41.25 | 13.08 |
| Kunming | 31.67 | 45.15 | 4.47 | 28.26 | 39.05 | -2.27 |
| Zhengzhou | 22.83 | 30.75 | 14.73 | 16.51 | 23.24 | -11.71 |
| Wuhan | 39.81 | 58.95 | 33.11 | 17.32 | 27.12 | -13.67 |
| Baoshan | 38.18 | 60.97 | 17.02 | 18.78 | 29.69 | -7.54 |
| Guangzhou | 26.78 | 39.79 | 12.37 | 14.22 | 21.85 | -6.14 |
| Sanya | 25.82 | 39.27 | 6.71 | 16.60 | 24.77 | -3.88 |
| Xianghe | 41.64 | 60.56 | 37.43 | 32.67 | 46.80 | -30.69 |

4.1.2 Parameter optimization of BRL

Compared to 17 first-class stations, Xianghe owns more available data which permits quantitative performance evaluation provided by tuned BRL. Here data samples at Xianghe are split into training and test sets. Training data set is used for nonlinear regression to determine empirical coefficients of tuned BRL (see Table 5), and test data set evaluates the performance improvement resulted from local tuning. Parameter optimization is implemented with R built-in function optim based on Nelder-Mead method (Nelder and Mead, 1965). Obviously, adjusted BRL can capture the spread and shape of ground data at Xianghe better (Fig. 3). For all-sky conditions, errors of DNI and DHI estimation decrease apparently, in which DNI rMAE, rRMSE and rMBE drop to 22.91%, 37.26% and 7.40%, DHI rMAE, rRMSE and rMBE reduce to 18.32%, 29.81% and -9.02%. Original BRL tends to overestimate DNI and underestimate DHI at Xianghe. After tuning, DNI rMBE drops from 37.43% to 7.40% and DHI rMBE decreases from -30.69% to -9.02%, which proves most of systematic bias is removed. For clear-sky ($k_t \geqslant 0.75$), cloudy

($0.3<k_t<0.75$) and overcast ($k_t \leq 0.3$) conditions, DNI and DHI errors correspondingly decrease in which the largest improvement occurs under overcast condition (Table 8).

Table 8. Error statistics of DNI and DHI modeling with BRL at Xianghe (unit: %). Original BRL and local-tuned BRL are sequentially validated with ground truth which spans from Jan 1st, 2007 to Dec 31st, 2011. Sky conditions are categorized to all sky, clear sky, cloudy sky and overcast sky based on clearness index $k_t$. Obviously, parameter optimization improves direct-diffuse separation performance of BRL.

|  |  | DNI | | | DHI | | |
| --- | --- | --- | --- | --- | --- | --- | --- |
| Model | Sky conditions | rMAE | rRMSE | rMBE | rMAE | rRMSE | rMBE |
| BRL | *All sky* | 41.64 | 60.56 | 37.43 | 32.67 | 46.80 | -30.69 |
|  | $k_t \geq 0.75$ | 30.33 | 42.10 | 29.64 | 59.19 | 80.27 | -57.84 |
|  | $0.3<k_t<0.75$ | 45.19 | 59.58 | 39.77 | 31.30 | 41.34 | -29.24 |
|  | $k_t \leq 0.3$ | 103.52 | 160.41 | 84.80 | 4.94 | 7.20 | -3.62 |
| Tuned BRL | *All sky* | 22.91 | 37.26 | 7.40 | 18.32 | 29.81 | -9.02 |
|  | $k_t \geq 0.75$ | 20.39 | 31.08 | 19.02 | 39.70 | 62.02 | -35.82 |
|  | $0.3<k_t<0.75$ | 23.54 | 33.19 | 3.00 | 16.71 | 24.45 | -6.49 |
|  | $k_t \leq 0.3$ | 41.99 | 110.91 | -6.78 | 3.63 | 5.41 | -0.12 |

4.2 Validation of GFS day-ahead solar irradiance forecasts

In Section 4.2.1, GFS day-ahead GHI, DNI and DHI forecasts are evaluated against ground truth at 18 stations. In Section 4.2.2, the influence of direct-diffuse separation on DNI and DHI forecasts are estimated with dataset at Xianghe.

4.2.1 GFS performance

For 18 locations, GHI forecasts rMAE and rRMSE are lower than DNI forecasts' which proves DNI is more difficult to be predicted than GHI by the model (Table 9). Generally, GHI forecasts rMBE is positive except Mohe, Harbin, Shenyang, and Beijing, implying that GFS overestimates global solar irradiance, which is consistent with the positive bias for GHI forecasts by GFS reported in Mathiesen and Kleissl (2011), Huang and Thatcher (2017). Similar to GHI forecasts, GFS generally over-predicts DNI in which rMBE is generally positive except Mohe, Shenyang, Beijing and Lhasa. A possible explanation for GHI and DNI over-prediction by GFS is anthropogenic aerosol. Most region in China is seriously affected by fine-mode industrial aerosols all year. Aerosol climatology in GFS is not enough to reflect the variability of aerosols produced by anthropogenic emissions over this region. Under clear-sky conditions, aerosol loading in the model is far lower than actual concentration in the atmosphere. Accordingly, scattering and absorption of solar irradiance by aerosols (i.e. aerosol direct effects) are underestimated. When clouds exist, column cloud droplet number concentration and liquid water path are increased by anthropogenic aerosols through aerosol indirect effects (Zhao et al., 2017). Consequently, for cloudy and overcast conditions, solar irradiance is further reduced.

Table 9. Error statistics of GFS day-ahead solar irradiance forecasts at 18 ground stations (unit: %). DNI and DHI are derived from original BRL. Solar irradiance measurements at 17 CMA first-class stations span from Jan 1$^{st}$, 2015 to Aug 31$^{st}$, 2016. Observed solar irradiance at Xianghe covers Jan 1$^{st}$, 2014 to Oct 31$^{st}$, 2015.

|         | GHI    |        |        | DNI    |        |        | DHI    |        |        |
|---------|--------|--------|--------|--------|--------|--------|--------|--------|--------|
| Station | rMAE   | rRMSE  | rMBE   | rMAE   | rRMSE  | rMBE   | rMAE   | rRMSE  | rMBE   |
| Mohe    | 51.29  | 68.94  | -11.45 | 128.42 | 168.98 | -8.13  | 67.28  | 88.42  | 7.72   |
| Harbin  | 51.10  | 65.89  | -7.77  | 141.29 | 192.06 | 4.04   | 62.45  | 80.33  | 9.76   |
| Urumqi  | 58.76  | 69.68  | 26.99  | 80.45  | 98.08  | 31.18  | 58.18  | 72.31  | 24.94  |
| Kashi   | 76.20  | 96.96  | 30.96  | 171.33 | 223.17 | 98.17  | 52.92  | 66.67  | -14.84 |
| Ejinaqi | 57.00  | 71.59  | 9.65   | 83.89  | 102.29 | 1.66   | 99.06  | 128.30 | 55.98  |
| Geermu  | 51.76  | 66.69  | 2.20   | 99.75  | 130.39 | 13.36  | 57.43  | 71.12  | -2.07  |
| Yuzhong | 61.10  | 78.57  | 15.92  | 133.13 | 177.29 | 55.66  | 56.44  | 73.09  | -0.31  |
| Shenyang| 55.05  | 67.04  | -17.07 | 109.05 | 136.64 | -44.60 | 71.09  | 92.62  | 28.24  |
| Beijing | 51.25  | 65.31  | -14.54 | 127.44 | 162.93 | -13.64 | 55.49  | 74.00  | -0.52  |
| Lhasa   | 56.40  | 70.71  | 2.40   | 106.47 | 134.96 | -1.28  | 63.62  | 79.04  | 8.38   |
| Chengdu | 128.94 | 156.22 | 92.76  | 328.54 | 457.42 | 270.64 | 94.96  | 116.29 | 56.41  |
| Kunming | 57.25  | 71.16  | 15.52  | 116.98 | 155.60 | 46.19  | 48.80  | 62.66  | -3.43  |
| Zhengzhou | 57.17 | 73.17 | 3.34 | 153.32 | 203.56 | 42.97 | 46.23 | 61.46 | -13.13 |
| Wuhan | 74.72 | 94.46 | 23.77 | 226.36 | 309.40 | 119.81 | 51.12 | 67.32 | -5.50 |
| Baoshan | 68.30 | 87.31 | 10.08 | 191.27 | 257.79 | 54.20 | 54.33 | 70.75 | 3.90 |
| Guangzhou | 74.15 | 94.64 | 24.75 | 191.50 | 264.76 | 91.83 | 51.06 | 67.48 | -1.81 |
| Sanya | 60.41 | 77.65 | 5.63 | 156.75 | 212.88 | 35.75 | 46.42 | 61.03 | -3.48 |
| Xianghe | 49.66 | 79.57 | 23.33 | 222.49 | 517.09 | 157.20 | 49.52 | 64.94 | -28.08 |

### 4.2.2 The influence of BRL tuning on DNI forecasts

Tuning the empirical coefficients of BRL significantly improves the accuracy for on-site DNI estimation (Section 4.1.2). Naturally, it is worthy to evaluate the influence of parameter optimization of BRL on DNI forecasts. Table 10 presents that adjusted BRL reduces DNI forecasts errors to a certain extent in comparison with original BRL only under cloudy and overcast conditions. The overall improvement is trivial for all-sky conditions which demonstrates the errors for DNI forecasts possibly stem from inherent physics of GFS model rather than direct-diffuse partition procedure. Hence, optimizing a direct-diffuse separation model with local observations may not be practical to increase on-site accuracy for DNI forecasts by NWP models.

Table 10. Error statistics of GFS day-ahead solar irradiance forecasts at Xianghe (unit: %). DNI and DHI forecasts are derived from GFS GHI forecasts via BRL. In order to evaluate the impact of parameter optimization of BRL on DNI and DHI forecasts, original BRL and tuned BRL are respectively employed in the calculation. Validation dataset span from Jan 1st, 2014 to Oct 31st, 2015. Sky conditions are categorized to all sky, clear sky, cloudy sky and overcast sky based on clearness index $k_t$. Tuned BRL fails to bring major improvements for DNI and DHI forecasts.

| | | GHI | | | DNI | | | DHI | | |
|---|---|---|---|---|---|---|---|---|---|---|
| Model | Sky conditions | rMAE | rRMSE | rMBE | rMAE | rRMSE | rMBE | rMAE | rRMSE | rMBE |
| | *All sky* | 49.66 | 79.57 | 23.33 | 222.49 | 517.09 | 157.20 | 49.52 | 64.94 | -28.08 |
| GFS-BRL | *$k_t \geq 0.75$* | 18.32 | 21.07 | -16.65 | 37.40 | 45.50 | -30.23 | 68.39 | 84.43 | 45.91 |
| | *$0.3 < k_t < 0.75$* | 42.76 | 69.94 | 17.51 | 231.09 | 526.88 | 166.66 | 46.35 | 59.64 | -40.00 |
| | *$k_t \leq 0.3$* | 183.00 | 284.51 | 164.92 | 15011.85 | 37309.79 | 14965.09 | 60.89 | 81.83 | 13.71 |
|  | *All sky* | 49.66 | 79.57 | 23.33 | 207.21 | 508.31 | 134.93 | 46.97 | 60.89 | -14.22 |
| GFS-TBRL | $k_t \geq 0.75$ | 18.32 | 21.07 | -16.65 | 39.64 | 47.95 | -34.59 | 81.05 | 99.40 | 66.46 |
|  | $0.3 < k_t < 0.75$ | 42.76 | 69.94 | 17.51 | 217.59 | 519.45 | 146.04 | 39.97 | 51.23 | -28.27 |
|  | $k_t \leq 0.3$ | 183.00 | 284.51 | 164.92 | 13027.18 | 36206.53 | 12969.10 | 75.12 | 102.89 | 37.53 |

4.3 Hybrid model evaluation

As long as Xianghe has longer observing period and research-class instruments, XGBoost-based hybrid model is trained, validated and tested with datasets at Xianghe. Section 4.3.1 describes the preprocessing of the inputs, model architecture, training procedure, and hyperparameter tuning etc. Section 4.3.2 presents evaluation results and importance of each input.

4.3.1 Model development

A hybrid model is designed to capture the nonlinear relationship between the inputs and the target in which the target is observed GHI, DNI and DHI in this application. In the context of machine learning, collinearity would occur and cause overfitting when two or more features (i.e. input variables) are correlated highly. To avoid collinearity, features should be filtered. In this work, potential features of XGBoost algorithm involve GFS variables, extraterrestrial horizontal irradiance and solar geometry parameters. For preprocessing the inputs, correlation coefficient is measured pairwise between two features, and then features with correlation coefficients greater than 0.9 are dropped. The remaining ones are kept as model inputs (Table 11). Afterward, dataset at Xianghe is split into training and test sets without introducing look-ahead bias in order to evaluate the fitted model's generalization ability on unseen data. 80% of available samples are used to train the model. The training dataset is further divided into *K* equally sized subsets without shuffling in order to select the optimal model parameters in which the procedure is known as *K*-fold cross validation. During model training, subsample ratio, regularization penalties, number of trees, minimal weight in a child, maximal tree depth, learning rate, minimal loss reduction, and subsample ratio are optimized through 6-fold cross-validated randomized search from a hyper-parameter distribution. Model fitting generates three models that correspond to GHI, DNI and DHI forecasting. In the end, 20% of the data samples are considered to be test dataset and is used to evaluate the overall performance of the final models.

Table 11. Feature inputs for hybrid model

| No. | Feature | Description | Unit |
| --- | --- | --- | --- |
| 1 | $k_t$ | hourly clearness index | unitless |
| 2 | $K_t$ | daily clearness index | unitless |
| 3 | AST | apparent solar time | hour |
| 4 | $sinα$ | sine of solar elevation angle | unitless |
| 5 | PRES | pressure | Pa |
| 6 | TMP | temperature | K |
| 7 | RH | relative humidity | % |
| 8 | SUNSD | sunshine duration | s |
| 9 | PWAT | precipitable water | kg/m$^2$ |
| 10 | TCDC | total cloud cover | % |

4.3.2 Model evaluation

The overall performance of GFS solar irradiance forecasts at Xianghe is enhanced with XGBoost significantly (Table 12). GHI forecasts rMAE, rRMSE and rMBE drop to 28.09%, 38.23% and -5.83% while corresponding error metrics for DHI forecasts decrease to 38.08%, 51.31% and 11.41%. DNI forecasts accuracy has the largest improvement in which rMAE, rRMSE and rMBE are reduced to 62.78%, 87.91% and -18.55%. These results suggest that this hybrid model for solar irradiance forecasts is far superior to direct NWP output by GFS model, and can be employed in operational forecast.

An advantage of XGBoost is that after gradient boosted decision trees are constructed it can estimate and rank each feature's importance score which indicates how valuable corresponding feature is in model training. In this paper, rank scores for features are retrieved from individual trained models for GHI, DNI and DHI forecasts (Fig. 4). A higher rank score of a feature when compared to another feature implies it is more critical for solar irradiance forecasts. For GHI forecasts, the sine of solar elevation angle $sinα$ and forecasted clearness index $k_t$ are the most important features. DNI forecasts are influenced much more by total cloud cover TCDC and relative humidity RH. For DHI forecasts, the sine of solar elevation angle $sin\alpha$ and temperature TMP play more important roles in model training. It should be kept in mind that the rankings for feature importance here merely reflect how valuable each input is to build gradient boosted decision trees at specific location, i.e., Xianghe.

Table 12. Error statistics of day-ahead solar irradiance forecasts based on GFS variables and XGBoost at Xianghe (unit: %). Observed solar irradiance at Xianghe covers Jan 1st, 2014 to Oct 31st, 2015.

|  | GHI | | | DNI | | | DHI | | |
| --- | --- | --- | --- | --- | --- | --- | --- | --- | --- |
| Model | rMAE | rRMSE | rMBE | rMAE | rRMSE | rMBE | rMAE | rRMSE | rMBE |
| GFS-XGBoost | 28.09 | 38.23 | -5.83 | 62.78 | 87.91 | -18.55 | 38.08 | 51.31 | 11.41 |
| GFS-BRL | 49.66 | 79.57 | 23.33 | 222.49 | 517.09 | 157.20 | 49.52 | 64.94 | -28.08 |

4.4 Comparison with benchmark forecasts

Lastly, GFS GHI forecasts at 18 stations and hybrid GHI forecasts at Xianghe are compared with results by persistence model. Table 13 presents that at 18 locations persistence forecasts rMAE and rRMSE are lower than GFS forecasts errors for all-sky conditions, indicating the accuracy of persistence model is better than GFS. Nevertheless, the performance of XGBoost-based model at Xianghe is much better than persistence forecasts, which implies this hybrid method has great potential as the best operational model at other sites over mainland.

Table 13. Error statistics of GHI persistence forecasts at 18 ground stations (unit: %). Solar irradiance measurements at 17 CMA first-class stations span from Jan 1st, 2015 to Aug 31st, 2016. Observed solar irradiance at Xianghe covers Jan 1st, 2014 to Oct 31st, 2015.

|  | GHI | | |
| --- | --- | --- | --- |
| Station | rMAE | rRMSE | rMBE |
| Mohe | 45.61 | 59.67 | 18.20 |
| Harbin | 44.35 | 56.48 | 25.58 |
| Urumqi | 30.29 | 38.72 | 5.72 |

| Station | | | |
|---|---|---|---|
| Kashi | 50.08 | 62.32 | 21.38 |
| Ejinaqi | 32.35 | 40.56 | 5.07 |
| Geermu | 36.00 | 43.69 | 0.24 |
| Yuzhong | 42.08 | 52.58 | 14.63 |
| Shenyang | 40.51 | 53.39 | 20.23 |
| Beijing | 33.64 | 44.42 | 8.88 |
| Lhasa | 40.54 | 48.46 | 3.00 |
| Chengdu | 108.51 | 125.30 | 97.58 |
| Kunming | 42.04 | 51.44 | 16.27 |
| Zhengzhou | 36.04 | 48.76 | 21.52 |
| Wuhan | 54.16 | 70.99 | 43.12 |
| Baoshan | 52.88 | 67.47 | 37.81 |
| Guangzhou | 57.10 | 74.98 | 46.98 |
| Sanya | 40.38 | 51.59 | 20.85 |
| Xianghe | 36.34 | 50.11 | -1.16 |

## 5. Conclusion

Day-ahead solar irradiance forecasts based on operational NWP model in China has yet to be characterized over the whole territory, compromising both the revenue of solar power plants and operation of TSOs throughout the country. To address this issue, GFS day-ahead irradiance forecasts are evaluated against hourly ground-based measurements at 18 high-quality radiometric stations in China in which direct and diffuse irradiance forecasts are calculated with direct-diffuse separation model BRL. Statistical indicators show that GHI and DNI forecasts are generally overestimated by GFS model, and DNI is much more difficult to predict reliably compared to GHI. Totally, GFS performance is not comparable to persistence model. Case study at Xianghe presents that although adjusted BRL model performs much better than the original one for both direct and diffuse irradiance estimation, unfortunately it is marginally important to improve DNI forecasts. Last, a hybrid model integrating GFS forecasts with XGBoost algorithm is built and verified against ground truth at Xianghe. It demonstrates the hybrid model's performance is much better than both crude output by GFS and persistence forecasts.

Given data availability, verification of MOS-corrected GFS forecasts is not performed in this paper. An idea for further work is to investigate the effects of bias correction resulted from MOS and other ML algorithms. Also, it is of interest to study the accuracy of solar irradiance forecasts made with other operational models such as ECMWF and JMA-GSM through model intercomparison.

**Declaration of Competing Interest**

None.

**Acknowledgments**

The authors would like to thank the National Natural Science Foundation of China for supporting this work (grant no. 41590875). The personnel of 17 first-class solar monitoring stations and BSRN Xianghe station used in this investigation are warmly thanked for establishing and maintaining this invaluable source of data.

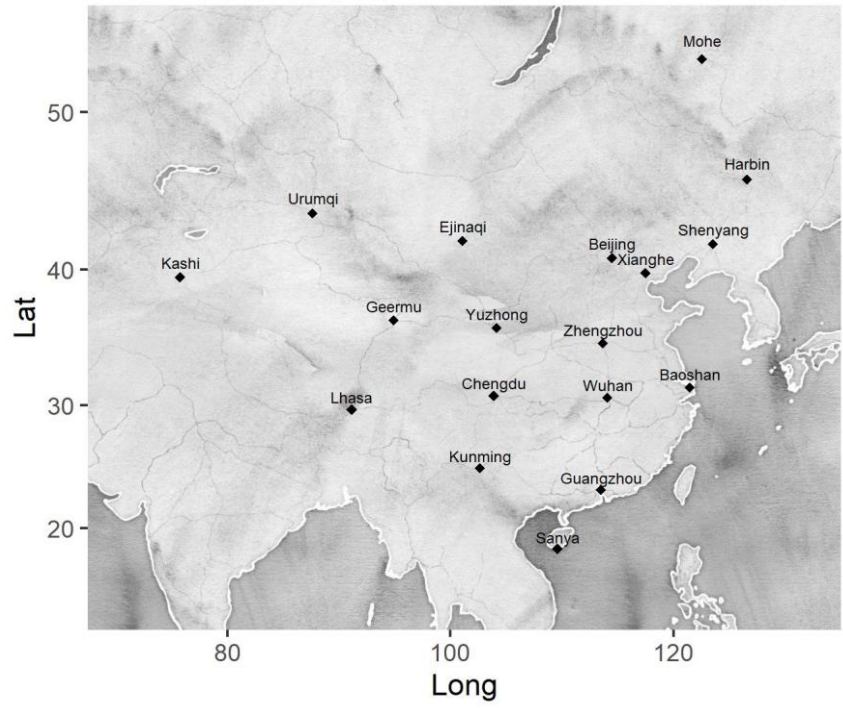

Fig. 1. Geographical distribution of the 18 stations

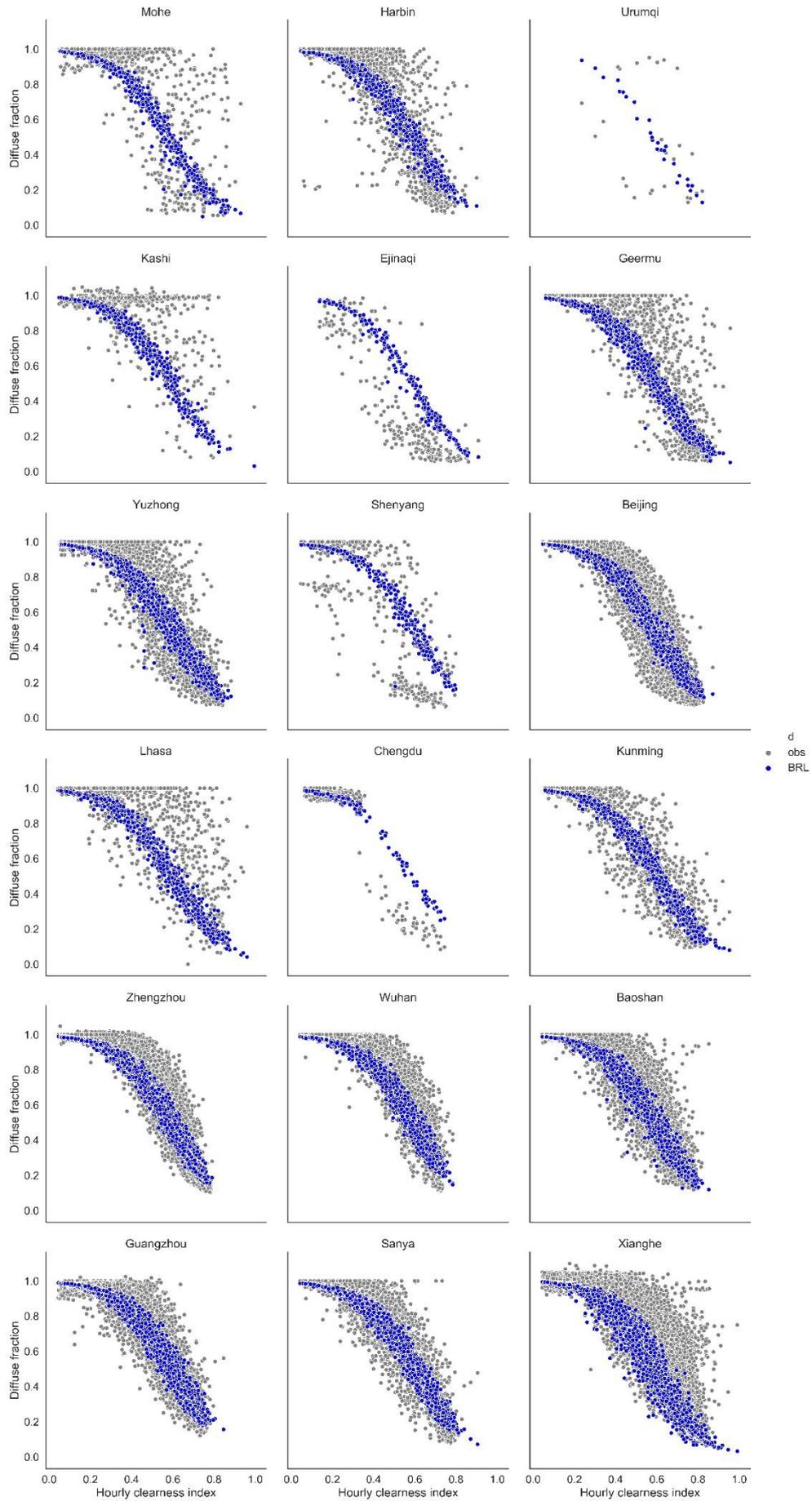

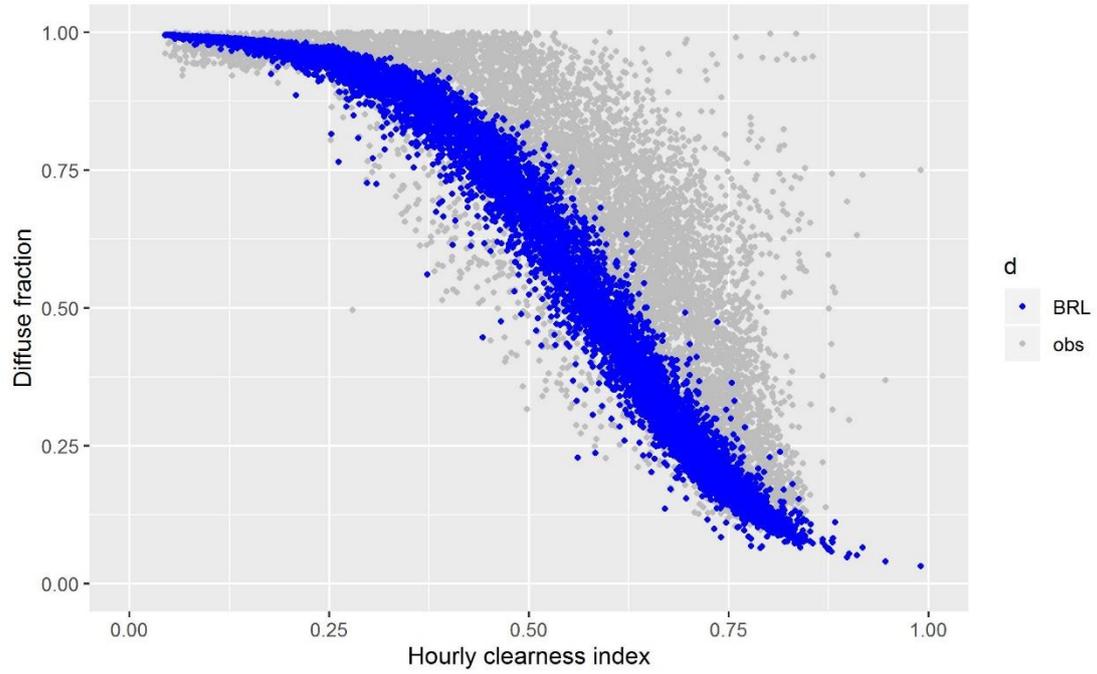

Fig. 2. Relationship between hourly clearness index $k_t$ and diffuse fraction $d$ at 18 stations. Grey points represent observed data, and model estimates with BRL are shown in blue.

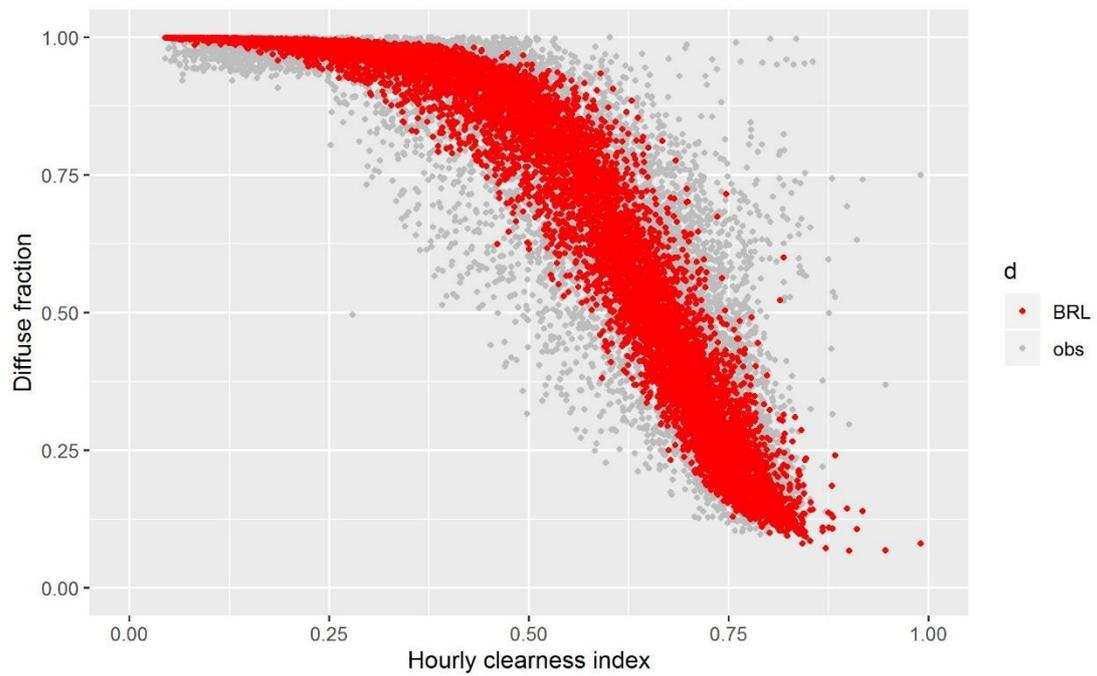

Fig. 3. Relationship between hourly clearness index $k_t$ and diffuse fraction $d$ at Xianghe. Grey points represent observed data, and model estimates with original BRL and tuned BRL are shown in blue and red.

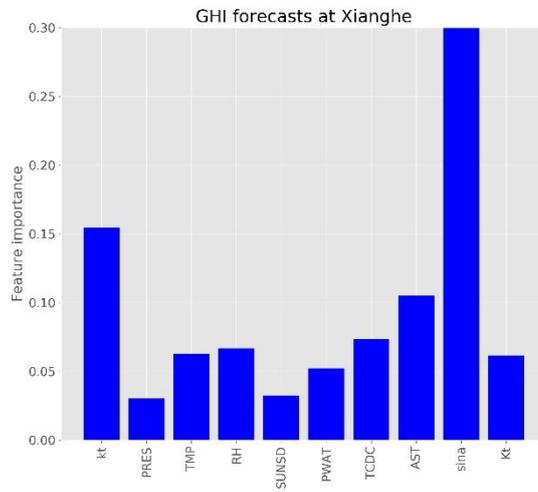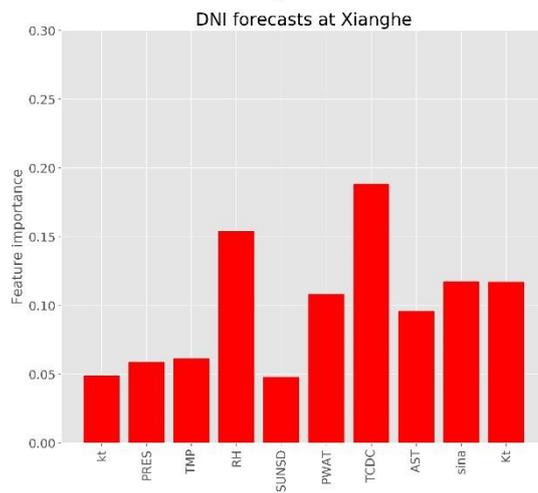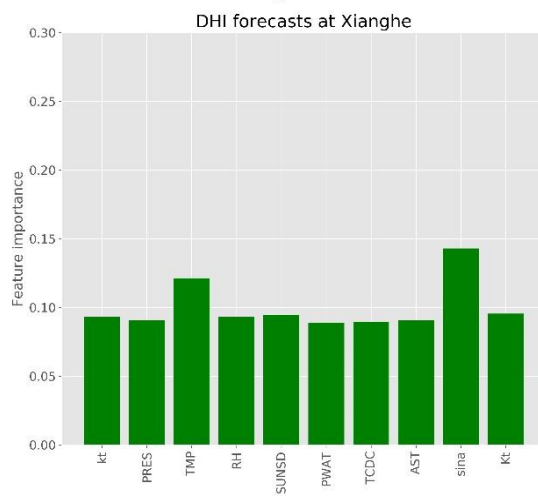

Fig. 4. Feature importance for XGBoost-based model. x-axis is features ordered by input indices. y-axis is importance scores for corresponding features. For GHI, *sinα* and forecasted clearness index $k_t$ are most important. For DNI, total cloud cover TCDC and relative humidity RH are key features. For DHI, *sinα* and temperature TMP are critical elements.